\documentclass{article}

\usepackage{PRIMEarxiv}

\usepackage[utf8]{inputenc} 
\usepackage[T1]{fontenc}    
\usepackage{hyperref}       
\usepackage{url}            
\usepackage{booktabs}       
\usepackage{amsfonts}       
\usepackage{nicefrac}       
\usepackage{microtype}      
\usepackage{lipsum}		
\usepackage{graphicx}
\usepackage{doi}

\usepackage{amsmath,amsthm,amsfonts,amssymb}
\usepackage{algorithm}
\usepackage{algpseudocode}
\usepackage{setspace}
\usepackage{booktabs}
\usepackage{listings}
\usepackage{color}

\title{Runtime-optimized Multi-way Stream Join Operator for Large-scale Streaming data}

\author
{Jinlong Hu *, Tingfeng Qiu \\
  School of Computer Science and Engineering \\
  South China University of Technology, Guangzhou, China\\
  \texttt{jlhu@scut.edu.cn} \\
}

\begin{document}
\maketitle

\begin{abstract}
Streaming computing enables the real-time processing of large volumes of data and offers significant advantages for various applications, including real-time recommendations, anomaly detection, and monitoring. The multi-way stream join operator facilitates the integration of multiple data streams into a single operator, allowing for a more comprehensive understanding by consolidating information from diverse sources. Although this operator is valuable in stream processing systems, its current probe order is determined prior to execution, making it challenging to adapt to real-time and unpredictable data streams, which can potentially diminish its operational efficiency. In this paper, we introduce a runtime-optimized multi-way stream join operator that incorporates various adaptive strategies to enhance the probe order during the joining of multi-way data streams. The operator's runtime operation is divided into cycles, during which relevant statistical information from the data streams is collected and updated. Historical statistical data is then utilized to predict the characteristics of the data streams in the current cycle using a quadratic exponential smoothing prediction method. An adaptive optimization algorithm based on a cost model, namely dpPick, is subsequently designed to refine the probe order, enabling better adaptation to real-time, unknown data streams and improving the operator's processing efficiency. Experiments conducted on the TPC-DS dataset demonstrate that the proposed multi-way stream join method significantly outperforms the comparative method in terms of processing efficiency.
\end{abstract}


\section{Introduction}
With the increasingly widespread application of real-time data streams—ranging from social media information flows to sensor data from the Internet of Things—the world is gradually entering an era focused on data streams in data processing. In this era, data streams are not only continuous and massive but also exhibit highly dynamic and uncertain content and structure. These characteristics present unprecedented challenges for the design and implementation of Data Stream Management Systems (DSMS) \cite{76,77}. When addressing resource-intensive operations such as multi-way stream joins, it becomes crucial to explore more efficient and flexible methods for handling and analyzing real-time data streams \cite{78,79,80}. In particular, multi-way stream join operators, which are essential technologies for implementing multi-way stream joins, hold significant practical applications and research potential for developing adaptive optimization strategies in real-time data streams.

Solutions have been proposed to optimize the probing order for multi-way stream join operators. Operators such as MJoin \cite{35} and GrubJoin \cite{19} employ low-selectivity-first probing order selection methods, while the MultiStream \cite{42} operator utilizes a greedy algorithm to enhance the probing order by prioritizing lower costs. These studies rely on static or known statistical information for probing order selection, which is appropriate for scenarios where the characteristics of data streams, such as those found in static databases, are well understood. However, in many practical applications, data streams are often generated in real-time, remain unknown, and continuously evolve. This dynamic nature means that static processing strategies may struggle to manage real-time data streams effectively, leading to a significant decline in data processing performance. To address this challenge, there is an urgent need for a method that can adaptively adjust probing orders during the runtime of data streams, ensuring optimal processing performance even in scenarios where data streams are unpredictable and constantly changing.

To address this issue, this paper proposes a runtime optimization-based multi-way stream join operator for state-based processing, incorporating a novel and comprehensive runtime adaptive optimization strategy into the multi-way data stream join process. Specifically, for state-based processing, a cost model that includes query costs is introduced. Based on this cost model, we propose an adaptive optimization algorithm for probing order, namely dpPick, along with an effective solution for the real-time acquisition and prediction of critical statistical information. Compared to previous research on probing order optimization, the main innovations are as follows: (1) Adaptive optimization of probing order during the runtime of multi-way stream operators, making the optimization strategy more suitable for unknown and changing data streams; (2) the introduction of query costs for states, which enhances the completeness and reasonableness of the cost model for multi-way stream join operations; and (3) the provision of a comprehensive and feasible strategy for collecting relevant statistical information, along with the proposal of secondary exponential smoothing prediction for critical statistical information, thereby improving the accuracy of cost calculations.

Through experiments and comparisons conducted on the TPC-DS dataset, the proposed method demonstrates superior processing efficiency compared to fixed initial order methods, MJoin, GrubJoin, and MultiStream. This adaptive optimization strategy enables the multi-way stream join operator to handle unknown data streams more flexibly and efficiently, thereby enhancing the performance of real-time data analysis.

\section{Related Work}
The multi-way stream join operation is a data processing technique used to combine data from multiple streams or sources based on shared attributes. This process involves matching or correlating data rows from different streams to create a unified data stream that incorporates information from various sources. Such join operations are frequently employed in data processing and analysis to integrate and synthesize information, facilitating deeper insights and informed decision-making. Given its high complexity and widespread application, the multi-way stream join is considered one of the most resource-intensive and crucial operations in stream processing, prompting some research into optimization strategies.

Studies have been conducted on load shedding in multi-way stream join processes, aiming to minimize the participation of irrelevant data in communication or computation by pre-filtering unjoinable tuples \cite{17,18}, selecting the most relevant data \cite{19}, and setting data priorities \cite{20}, among other strategies, to maximize output rates while utilizing the same resources. In terms of load balancing in multi-way stream joins, Wang et al. \cite{21} sliced and allocated states based on a cost model of query response, dynamically relocating states at runtime for adaptive load balancing. Fan et al. \cite{22} proposed a GroJoin optimization algorithm based on the micro-batch processing model to address data skewness issues in multi-way stream joins. Regarding multi-query optimization in multi-way stream joins, researchers have explored the reuse of concepts such as shared operators and shared data to reduce redundant computation and storage. Jonathan et al. \cite{23} introduced an incremental sharing execution operator method based on wide-area flow analysis, while Karimov et al. developed a system called AStream \cite{24} for sharing multiple stream query resources and subsequently introduced the first shared real-time stream processing engine, AJoin \cite{25}. Le-Phuoc et al. \cite{26} proposed connecting multiple streams through shared windows and result routing, while Dossinger et al. \cite{27} formulated an integer linear programming (ILP) model for selecting partitions and routing tuples to minimize detection load.

The join order in multi-way stream joins is a crucial factor influencing operational efficiency. The size of the intermediate state in binary join trees is closely related to the structure of these join trees. Factors such as the selectivity of each join condition \cite{28} and the input rate \cite{29} significantly impact the intermediate state size of the entire join tree. Additionally, the quality of the probing order for multi-way stream join operators directly affects the amount of redundant computation. Extensive research has been conducted on optimization problems related to join order.

In the domain of batch processing, Vance et al. \cite{30} proposed a Cartesian product optimizer that employs dynamic programming algorithms to explore the planning space and calculate costs. They combined this approach with estimates of intermediate result cardinality to develop a join order optimizer. Wi et al. \cite{31} addressed the enumeration problem of binary join trees by proposing the use of multi-way join units and introduced a mechanism for the bottom-up expansion of small join units into larger ones based on cost considerations. Ji et al. \cite{32} designed and implemented a join operator tree, called joinTree, on Flink to facilitate multi-source data joins. To mitigate computational complexity in multi-source data join operations, they proposed a heuristic optimization strategy that includes a cost model and an ant colony algorithm to identify the optimal join sequence. Additionally, they introduced a data compression-based repartitioning strategy to minimize data communication between computational nodes. Lai et al. \cite{34} accelerated multi-way joins using GPUs. In contrast to batch data, streaming data is inherently more unpredictable and requires real-time processing, rendering these batch-related strategies inadequate for addressing challenges in stream processing.

Handling multiple input streams with highly variable and unpredictable rates presents a significant challenge. Traditional relational query optimizers depend on table cardinality to estimate query plan costs; however, for streams with infinite arrival rates, cardinality information remains unknown. To address this issue, Viglas et al. \cite{29} proposed a shift from conventional cardinality methods to rate-based approaches and introduced an optimization framework aimed at maximizing the output rates of query plans.

Research has been conducted on optimization techniques based on binary join trees. Gomes et al. \cite{15} proposed a dynamic programming algorithm, OptDP, for continuous multi-join queries on data streams utilizing sliding windows. They introduced a polynomial-time greedy algorithm, XGreedyJoin, which considers factors such as arrival rate, window size, and join rate to derive the most optimal join tree, thereby maximizing the throughput of multi-way stream joins within sliding windows. Following a change in the join plan, intermediate result migration is executed through state-moving strategies, including node reuse and result reuse. Lee et al. \cite{16} developed a greedy-based k-EGA algorithm for continuous multi-way join queries on data streams, with the goal of generating optimized query plans. By concurrently tracking a set of potential plans based on cost statistics tables, they adeptly managed the number of plans monitored to strike a balance between plan generation time and optimality. Alterations in the structure of binary join trees necessitate corresponding migrations of intermediate states, which can be complex and costly, thereby limiting their flexibility in dynamic data streams.

To address the issue of intermediate state generation in binary join trees, Viglas et al. proposed the multi-way stream join operator, MJoin \cite{35}, which facilitates multi-way stream joins within a single operator through iterative probing. They adopted a low selectivity priority strategy for selecting the probing order. Experiments demonstrated that implementing multi-way stream joins within a single operator is a significant advancement. Gedik et al. \cite{19} introduced the GrubJoin operator, which also employs the low selectivity priority strategy from the MJoin operator for join order selection. To minimize inter-node communication costs, Zhou et al. \cite{36} proposed a partition-based multi-way join scheme, PMJoin, derived from the MJoin operator. They partitioned data streams and explored various partition strategies, including rate-based, hash-based, and random approaches. Different partitioned substreams execute distinct join plans, and a heuristic optimization algorithm was developed based on the rate-based cost model to enhance these join plans. BaBu et al. \cite{37} introduced an adaptive greedy algorithm, A-Greedy, for the adaptive sorting of pipeline filters, which adjusts to achieve improved sorting by monitoring and recording filter selectivity. The probing operation serves as an instance of a filter, and the A-Greedy algorithm can be extended to the adaptive sorting of multi-way stream joins, as implemented in the StreaMon \cite{38} engine. Golab et al. \cite{39} investigated incremental multi-way join algorithms on sliding windows of data streams, proposing incremental multi-way joins in a nested loop fashion. They also introduced a heuristic join sorting algorithm based on unit time cost models to identify an effective join order without exhaustively searching the entire space. Le-Phuoc et al. \cite{26} modeled join operations on RDF streams as continuous multi-way join operations and presented a recursive cost model. They proposed both a one-step adaptive greedy algorithm and a two-step adaptive optimization algorithm that considers shortcut conditions, necessitating the maintenance of detection counts for all keys across each stream, which incurs significant memory overhead during long-term state processing. Dossinger et al., building on BiStream \cite{40}, extended the multi-way stream join operator to MultiStream \cite{41,42}, which similarly implements multi-way joins through iterative probing and proposes a greedy strategy for optimizing the probing sequence based on probing costs. Yu et al. introduced an efficient three-way distributed stream join system, TriJoin \cite{43}, which symmetrically partitions tuples and reuses intermediate results to reduce processing latency.

Although some research has been conducted on optimizing the probing order of multi-way stream join operators, existing studies predominantly rely on known statistical information about data streams for cost calculations. This approach—optimizing the probing order prior to operator execution—is effective for data streams originating from static databases but poses challenges when applied to real-time data streams that undergo unpredictable changes. Additionally, the cost modeling in existing research primarily focuses on window mechanisms, often neglecting the query cost associated with state. Consequently, the runtime optimization of state-based multi-way stream join operators in the context of real-time data streams remains a complex and exploratory area of research.

\section{Runtime-Optimized Multi-way Stream Join Operator}
\subsection{Overview of Multi-way Stream Join Operator}
\begin{figure}[ht]
  \centering
  \includegraphics[width=0.7\linewidth]{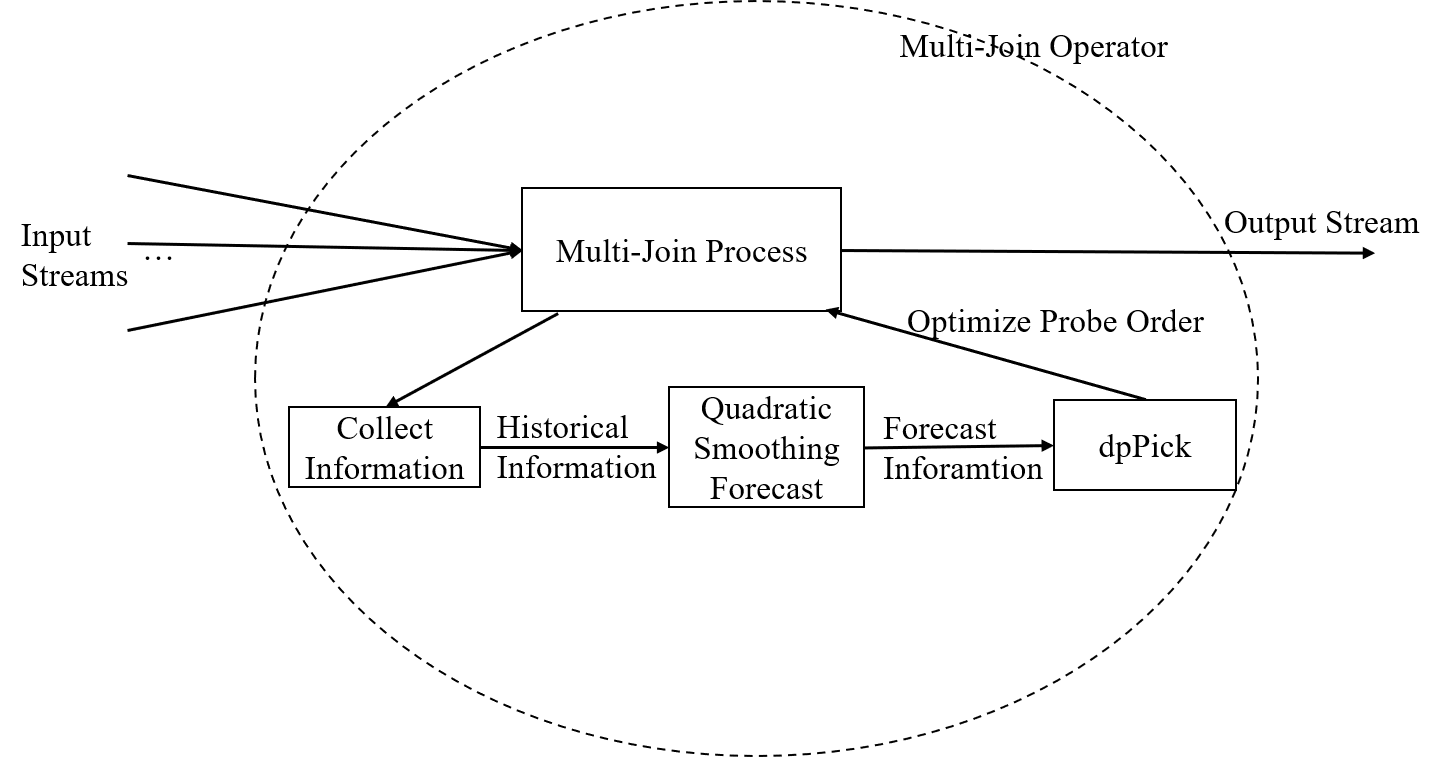}
  \caption{Overview of Multi-way Stream Join Operator}
  \label{fig:operator}
\end{figure}

This section introduces the proposed multi-way stream join operator, as illustrated in Fig.~\ref{fig:operator}. When executing the join operation on multiple data streams, the input streams are directed into the multi-way stream join operator, which processes them and produces the joined result data stream. The operator utilizes an iterative probing approach \cite{42} for multi-way stream joining. Its runtime is divided into cycles, during which it continuously gathers relevant statistical information for cost estimation from the data streams. Based on a timer cycle, the operator optimizes the probing order by leveraging historical cycle statistics to perform secondary exponential smoothing predictions, thereby generating predictive information for the current runtime cycle. This predictive information is incorporated into the probing order optimization algorithm (dpPick), which is developed based on the cost model, resulting in an updated and optimized probing order. This process facilitates adaptive runtime optimization for unknown data streams, significantly enhancing the efficiency of the operator's runtime.

 The multi-way stream joining process of the state-based multi-way stream join operator is detailed in Algorithm~\ref{multijoin}. Its input consists of a set of state backends \(B[1 \ldots n]\), each storing data from an input stream, along with a corresponding set of probing orders \(O[1 \ldots n]\) that specify the probing order for data stream joins, and an optimization time period \(T\), which defines the interval for optimization operations. The algorithm begins by fetching the current time (line 1) and calculating the next time to invoke optimization operations (line 2), ensuring periodic optimization of probing orders within the designated cycle time \( T \). Subsequently, the algorithm enters an infinite loop (lines 3-16), continuously detecting and processing newly arrived tuples, executing periodic optimization tasks, and performing asynchronous data cleaning. The algorithm asynchronously removes expired data from all state backends (line 4) to prevent state inflation and maintain data freshness. This is accomplished by invoking the \(\text{removeExpiredData()}\) procedure for each state backend \( B[i] \), minimizing interference with the main processing flow through asynchronous operations, thereby preserving high operator performance. The algorithm then checks whether the current time has reached or exceeded the scheduled time for the next optimization operation (line 6). If it has, the algorithm optimizes the probing orders \(O\) for each input stream by invoking the \(\text{optimizeProbeOrder\_dpPick()}\) function (line 7). This process dynamically adjusts the probing order based on the characteristics of the current data stream and the efficiency of previous probing attempts. The optimized probing order is designated as the scheduled probing order for the next runtime cycle, with the goal of enhancing both the efficiency and accuracy of data processing. Subsequently, the next optimization time point is updated (line 10), ensuring a dynamic adjustment of join orders based on the latest state of data streams to enhance the efficiency of join operations. Finally, when a new tuple \( t \) arrives at any input stream \( i \) (line 12), the algorithm performs an insertion operation (line 13), placing the tuple into the corresponding state backend and immediately triggering the iterative probing process, as indicated in line 14, by promptly invoking the \(\text{probeByOrder()}\) method, which iteratively probes according to the predefined probing order. This approach enables the operator to gradually establish joins between different data sources, laying the groundwork for final data integration and analysis. The efficiency of the probing process directly affects the operator's ability to process data streams, making the optimization of the probing order a crucial factor in enhancing operator performance.

\begin{algorithm}[ht]
\caption{MultiJoin}
\label{multijoin}
\begin{algorithmic}[1]
\State $currentTime \gets \text{getCurrentTime()}$
\State $nextCallTime \gets currentTime + T$
\While{true}
    \State asynchronously $\text{removeExpiredData}(B[1 \ldots n])$     
    \State $currentTime \gets \text{getCurrentTime()}$
    \If{$currentTime \geq nextCallTime$}     
        \State $O \gets \text{optimizeProbeOrder\_dpPick}(O)$     
        \State $nextCallTime \gets currentTime + T$
    \EndIf
    \If{a new tuple $t$ arrives at stream $i$}
        \State $B[i].\text{insert}(t)$     
        \State $\text{probeByOrder}(t,O[i])$     
    \EndIf
\EndWhile
\end{algorithmic}
\end{algorithm}

Algorithm~\ref{probe} elaborates on the specific process of the $\text{probeByOrder()}$ algorithm, which iteratively probes according to a predefined probing order. The algorithm takes as input the tuple $t$ to be probed and the predefined probing order $\{< {l_1},{r_1} > , \ldots , < {l_n},{r_n} > \}$. Starting from line 1, it queries the list of tuples that match tuple $t$. If the queried list is not empty (line 2), the $\text{match()}$ function is employed to generate a set of matching tuples, denoted as $matched$, and the iteration proceeds to the next probing step (lines 3-4) until no more matching tuples can be found or all predefined probing orders have been exhausted. Once a probing step fails to identify any matching tuples, the algorithm halts further probing (line 6), as there are no additional join opportunities along the current path. This mechanism prevents unnecessary redundant computations, thereby enhancing processing performance for the operator, particularly when managing large-scale data streams. Through this approach, the $\text{probeByOrder()}$ algorithm effectively directs the execution of the multi-way stream join operator by utilizing predefined probing orders and dynamically optimizing probing strategies, ensuring efficient and accurate data processing. The success of this method is attributed to the careful design and real-time adjustment of join orders, which not only improves query efficiency but also significantly reduces resource consumption, making it particularly suitable for handling large-scale and high-complexity data stream join scenarios.

\begin{algorithm}
\caption{ProbeByOrder}
\label{probe}
\begin{algorithmic}[1]
\State $queryedList \gets B[r_1].\text{query}(t)$
\If{$\text{size}(queryedList) > 0$}
\State $matched \gets \text{match}(t,queryedList)$
\State $\text{probeByOrder}(matched,{ < l_1,r_1 > , \ldots , < l_n,r_n > } \setminus { < l_1,r_1 > })$
\Else
\State \textbf{break}
\EndIf
\end{algorithmic}
\end{algorithm}

\subsection{Cost Model for Optimizing Multi-way Join}

In the field of real-time data stream processing, data stream sources may experience fluctuations due to varying data arrival rates, with many of these fluctuations arising from unpredictable, slow, or bursty network traffic. Given the streaming nature of input data, the characteristics of data streams are often unknown and subject to dynamic changes, leading to uncertainty in the data distribution itself. Consequently, establishing models for operator selectivity becomes challenging, complicating the selection of an efficient probing order for operators. To address the continually changing selectivity of join predicates in real-time data streams, this paper presents a runtime adaptive optimization method for state-based multi-way stream join operators, allowing the operator to identify a more effective execution plan during runtime. As demonstrated in Algorithm 3-1, this paper utilizes the $\text{optimizeProbeOrder\_dpPick()}$ algorithm, referred to as the dpPick algorithm, to dynamically determine the optimal probing order for multi-way join processes involving newly arriving tuples in each runtime cycle.

In line with this approach, this paper proposes a time cost model for state-based multi-way stream join operators, aiming to establish evaluation criteria for selecting the optimal probing order for each runtime cycle. To accurately estimate the total time cost incurred by probing operations between data streams, this model abstracts the probing operations as a series of paired data stream pairs, with each pair represented by a probing pair $\langle {l_i},{r_i}\rangle$, corresponding to a specific probing operation ${o_i}$. Here, ${l_i}$ and ${r_i}$ denote the input stream numbers of the probing end and the probed end, respectively. These operations form an ordered set ${o_1},{o_2}, \ldots ,{o_n}$, representing a non-circular join sequence created by $n$ inputs, which consists of $n - 1$ probing operations and the termination operation ${o_n}$. Each input stream corresponds to such an ordered set. Considering the recursive nature of the iterative probing process, its cost function is defined recursively in Eq.~\ref{eq:c1}.

\begin{equation}
{C_{{o_i}}} = {C_{{q_{{r_i}}}}} + \gamma _{{r_i}}^{{l_i}} \cdot ({C_{m_{{r_i}}^{{l_i}}}} + {C_{{o_{i + 1}}}})
\label{eq:c1}
\end{equation}
Eq.~\ref{eq:c1} provides an estimation of the overall expected total cost associated with the probing sequence leading to ${o_n}$. Here, ${C_{{o_i}}}$ represents the accumulated expected total cost starting from the $i$-th probing action, while ${C_{{q_{{r_i}}}}}$ denotes the expected cost of executing a query on the ${r_i}$-th state backend. The term $\gamma _{{r_i}}^{{l_i}}$ indicates the probability of successfully querying data from the ${l_i}$-th data stream on the ${r_i}$-th state backend, and ${C_{m_{{r_i}}^{{l_i}}}}$ represents the expected cost of matching tuples from the ${l_i}$-th data stream on the ${r_i}$-th state backend. Additionally, ${C_{{o_{i + 1}}}}$ signifies the accumulated expected total cost starting from subsequent probing actions. Both cost terms, ${C_{m_{{r_i}}^{{l_i}}}}$ and ${C_{{o_{i + 1}}}}$, are incurred only when the data query is successful; therefore, they are multiplied by the probability of a successful query, which is the likelihood of performing matching operations. For the termination operation ${o_n}$ in the join sequence, which signifies the end of the probing sequence, its cost is defined as zero, as shown in Eq.~\ref{eq:c2}.

\begin{equation}
{C_{{o_n}}} = 0
\label{eq:c2}
\end{equation}
Thus, the total cost $C$ of the entire probing sequence can be determined by the recursive formula for initiating the first join operation ${o_1}$, as shown in Eq.~\ref{eq:c3}.
\begin{equation}
C = {C_{{o_1}}}
\label{eq:c3}
\end{equation}
However, to accurately calculate the value of this cost function, it is essential to collect the execution time for each query operation and matching operation, and to iteratively compute all potential probing sequences separately. In practical application scenarios, the system load and computational costs associated with this approach would be substantial. To address this challenge, this paper proposes an approximation estimation method that utilizes relevant statistical information to estimate the cost and identify an approximation of the optimal probing sequence among all possible probing orders.

\section{Adaptive Optimization}
\subsection{Cost Estimation}

When constructing the cost model, both the costs of query and matching operations are taken into account, and the probability of query success is integrated into the cost estimation, resulting in a comprehensive cost evaluation framework. The costs of the query operation, denoted as ${C_{{q_{{r_i}}}}}$, and the matching operation, represented as ${C_{m_{{r_i}}^{{l_i}}}}$, are approximately defined by Eq.~\ref{eq:c4} and Eq.~\ref{eq:c7}, respectively. The cost of the query operation, ${C_{{q_{{r_i}}}}}$, reflects the resource consumption incurred by executing a query on the data stream backend ${r_i}$, as illustrated in Eq.~\ref{eq:c4}. To quantify this consumption, a query cost constant coefficient, ${\alpha _q}$, is introduced, which is assumed to be proportional to a function $f({\kappa _{{r_i}}})$ that reflects the influence of the number of keys. Here, ${\kappa _{{r_i}}}$ denotes the number of keys in backend ${r_i}$, and the function $f({\kappa _{{r_i}}})$ describes how the number of keys affects the resource consumption of the query engine. The form of this function depends on the design of the backend's data structure and the actual impact of the number of keys on performance, illustrating the increased resource demand for computing indexes and performing matching operations as the number of keys rises.

\begin{equation}
\label{eq:c4}
{C_{{q_{{r_i}}}}} \approx {\alpha _q} \cdot f({\kappa _{{r_i}}})
\end{equation}

In this paper, a specific definition is provided for the function \( f(\kappa) \). When the backend data structure is a hash table, \( f(\kappa) \) can be defined as shown in Equation \ref{eq:c5}, where \( \kappa \) denotes the number of key-value pairs in the hash table, and \( m \) represents the number of slots available in the hash table. The variable \( c \) indicates the basic query cost, which includes operations such as computing hash values and accessing array indices; these are fixed costs incurred by each query operation. The term \( \frac{\kappa}{2m} \) accounts for the additional cost resulting from key collisions, which is reflected in the average length of the linked list traversal when conflicts occur. During query operations, it may be necessary to traverse the linked list to locate the matching key when multiple keys are mapped to the same index.

\begin{equation}
\label{eq:c5}
f(\kappa ) = c + \frac{\kappa }{{2m}}
\end{equation}

When the backend's data structure is designed as an ordered array, \( f(\kappa) \) can be defined as shown in Eq.~\ref{eq:c6}. The query process utilizes binary search, resulting in a query cost that is logarithmically related to the number of keys.

\begin{equation}
\label{eq:c6}
f(\kappa ) = \log (\kappa )
\end{equation}

The cost of the matching operation, denoted as ${C_{m_{{r_i}}^{{l_i}}}}$, reflects the resource consumption incurred during the matching process in the data stream. As illustrated in Eq.~\ref{eq:c7}, the estimated time cost of the matching process is approximately calculated by multiplying the matching cost time coefficient, ${\alpha _m}$, by the corresponding average matching number, $\mu _{{r_i}}^{{l_i}}$.

\begin{equation}
\label{eq:c7}
{C_{m_{{r_i}}^{{l_i}}}} \approx {\alpha _m} \cdot \mu _{{r_i}}^{{l_i}}
\end{equation}

Finally, by synthesizing the definition of the cost model, this paper proposes a formula for calculating the total cost of probing operations, specifically Eq.~\ref{eq:c8}. This formula incorporates the query cost, matching cost, and the anticipated total cost of subsequent probing operations, providing an approximate estimation of the expected total cost from the current $i$-th probing operation to the conclusion of the probing sequence.

\begin{equation}
\label{eq:c8}
{C_{{o_i}}} \approx {\alpha _q} \cdot f({\kappa _{{r_i}}}) + \gamma _{{r_i}}^{{l_i}} \cdot ({\alpha _m} \cdot \mu _{{r_i}}^{{l_i}} + {C_{{o_{i + 1}}}})
\end{equation}

The significance of the aforementioned approximation estimation stems from the dynamic and uncertain nature of data in real-time data processing environments, which renders precise cost calculations exceedingly complex and computationally intensive. By employing these approximation methods, operators can efficiently estimate the costs associated with probing operations, thereby maintaining computational efficiency. This approach provides valuable evaluation criteria for optimizing the probing order in multi-way stream join processing.

\subsection{Probe Order Optimization}
To effectively optimize the probe order of operators, this paper proposes a probe order optimization algorithm (\texttt{dpPick}) based on a cost model, as detailed in Algorithm~\ref{dppick}. The objective of this algorithm is to select an optimal probe sequence for each input stream in order to minimize the total cost of probe operations. The core concept of this algorithm involves generating all possible probe orders based on the join graph, utilizing the cost model and cost estimation formulas, and selecting the probe sequence with the lowest potential cost from all available options. To reduce time complexity, the algorithm recursively calculates and memoizes the costs of probe sequences of varying lengths, ultimately selecting the one with the minimum cost estimation among all possible probe orders to replace the original probe sequence, thereby achieving probe order optimization.

The process of optimizing the probe order based on the cost model is detailed in Algorithm~\ref{dppick}. In the algorithm's input, $\kappa$ represents the number of keys in the state backends of each stream, where $\kappa[i]$ denotes the number of keys in the state backend of stream $i$. ${\alpha _q}$ is the cost coefficient for query operations, while ${\alpha _m}$ is the cost coefficient for match operations. The variable $\mu$ represents an array of average match counts, with $\mu[i][j]$ indicating the average match count when stream $i$ matches stream $j$. Additionally, $\gamma$ represents an array of match success probabilities, where $\gamma[i][j]$ indicates the probability of stream $i$ successfully matching stream $j$. Finally, $G$ denotes the join graph, in which each edge represents a potential probe pair.

The algorithm begins by initializing two key dictionaries (lines 1-2): `costDict`, an empty dictionary used to store the optimal probe sequence and its corresponding cost for each stream, and `subMemo`, a dictionary that contains the costs of empty sequences. The critical role of `subMemo` is to record the costs of previously calculated subsequences, enabling efficient memoization. Here, `subMemo` is initialized with a mapping of zero cost for empty sequences, representing the state before any probe operations have been conducted. Next, the algorithm iterates over all possible probe sequences and calculates their costs (lines 15-24), utilizing the `getSequence()` function to retrieve all possible complete probe sequences from the graph \( G \). For each sequence, it extracts the first probe pair \((i, j)\) along with its subsequent subsequence and calculates the total cost of this sequence using the `calculateCost()` function. This process recursively handles the cost calculation of subsequences and relies on `subMemo` to avoid redundant calculations. Based on the calculated total cost, it updates `costDict` with the optimal probe sequence and cost for the corresponding stream, only making updates when the newly calculated cost is lower than the previously recorded minimum cost. 

The recursive function \texttt{calculateCost()} (lines 3-14) encapsulates the core logic of the algorithm. It computes the cost of the current probe pair and its subsequences while determining whether to recalculate the cost of the subsequence $subCost$ based on $subMemo$. This approach ensures that the cost of each identical subsequence is calculated only once during the algorithm's entire runtime, significantly enhancing its efficiency. The function estimates costs using statistical information, calculating the query cost $queryCost$ and match cost $matchCost$ separately (lines 4-5). If the cost of the subsequence is already present in $subMemo$, it directly utilizes this cost value to avoid redundant calculations (lines 6-7). If the cost of the subsequence has not yet been calculated, the \texttt{calculateCost()} method recursively invokes itself to compute this cost and stores the result in $subMemo$ (lines 8-10). Subsequently, based on the cost model, it calculates the total cost of the current probe sequence (line 12). Finally, when the algorithm concludes, it returns $costDict$, which contains the optimal probe sequence and its corresponding cost for each stream. Throughout the entire algorithmic process, the implementation of updating and memoization strategies for probe sequences and the costs of all data streams helps to eliminate many redundant calculations of subsequence costs, thereby improving the algorithm's efficiency.

\begin{algorithm}
\caption{optimizeProbeOrder\_dpPick}
\label{dppick}
\begin{algorithmic}[1]
\Require $\kappa ,{\alpha _q},{\alpha _m},\mu ,\gamma ,G$   
\Ensure $\text{costDict}$       

\State $costDict \gets \{ \} $       
\State $\text{subMemo} \gets \{ (\text{null},0)\} $      

\Function{calculateCost}{$(i,j),\text{subSequence}$}   
\State $matchCost \gets {\alpha _m} \cdot \mu [i][j]$ 
\State $queryCost \gets {\alpha _q} \cdot f(\kappa [j])$ 
\If{$\text{subSequence} \in \text{subMemo}$}
    \State $subCost \gets \text{subMemo[subSequence]}$ 
\Else
    \State $subCost \gets \text{calculateCost}(subSequence[0],subSequence[1:])$ 
    \State $\text{subMemo[subsequence]} \gets \text{subCost}$    
\EndIf
\State $cost \gets queryCost + \gamma [i][j] \cdot (matchCost + \text{subCost})$   
\State \Return $\text{cost}$ 
\EndFunction

\State $allPossibleSequences \gets \text{getSequences}(G)$     
\ForAll{$\text{possibleSequence}$ in $\text{allPossibleSequences}$}
    \State $(i,j) \gets \text{possibleSequence[0]}$ 
    \State $\text{subSequence} \gets \text{possibleSequence[1:]}$ 
    \State $cost \gets \text{calculateCost}((i,j),\text{subSequence})$ 
    \State $minCost \gets \text{costDict.getOrDefault}(i,(\text{null},\infty ))[1]$
    \If{$\text{cost} < \text{minCost}$}
        \State $costDict[i] \gets (\text{possibleSequence},\text{cost})$     
    \EndIf
\EndFor
\State \Return $\text{costDict}$
\end{algorithmic}
\end{algorithm}

Algorithm ~\ref{getSequence} outlines the operational principles of the \texttt{getSequences()} method. **Input:** a graph \( G(V, E) \), where \( V \) denotes the set of nodes and \( E \) signifies the set of edges. **Output:** all possible probe orders \( \text{allSequences} \), with each order representing a permutation of probe pairs that constitutes a potential probe path. Initialization is performed first (line 15). An empty list, \( \text{allSequences} \), is created to store all probe orders. The depth-first search (DFS) function, dfs(), is then initiated for each join node (lines 16-20). The algorithm iterates over each node \( V \) in the graph \( G \). For each node, it is marked as visited (added to the \( \text{visited} \) set), and a depth-first search is initiated from that node. The dfs() function executes the depth-first search (lines 1-14). If the size of the \( \text{visited} \) set equals the size of the node set \( V \), it indicates that all nodes have been visited, and the current sequence encompasses all nodes in the graph. At this point, the current sequence is copied to \( \text{allSequences} \) and then returned (lines 2-4). The function traverses all edges in the graph \( G \) (line 5). For each edge, if the starting point has been visited and the endpoint has not yet been visited, the following operations are performed (line 6): the endpoint is marked as visited (added to the \( \text{visited} \) set), and this edge is added to the current probe sequence \( \text{currentSequence} \) (lines 7-8). The dfs() function is then recursively called with the current \( \text{visited} \) set as the new starting point to continue exploration (line 9). After the recursive call returns, the algorithm backtracks, undoing the modifications to the \( \text{visited} \) set and \( \text{currentSequence} \) to explore other possible paths (lines 10-11). Finally, all probe sequences in \( \text{allSequences} \) are returned (line 21). Through recursive depth-first search, the algorithm explores all possible paths starting from each node in the graph, ensuring that the generated probe sequences are comprehensive.

\begin{algorithm}
\caption{getSequences}
\label{getSequence}
\begin{algorithmic}[1]
\Require $G(V,E)$           
\Ensure $allSequences$  
\Function{dfs}{$visited,currentSequence,allSequences,G$} 
    \If{$\text{visited.size}$ equals $|G.\text{nodes}|$}
        \State $allSequences.\text{add}(currentSequence)$ \Return
    \EndIf 
    \ForAll{$edge$ in $G.\text{edges}$}
        \If{$\text{visited.contains}(edge.\text{startNode})$ and not $\text{visited.contains}(edge.\text{endNode})$}
            \State $\text{visited.add}(edge.\text{endNode})$
            \State $\text{currentSequence.add}(edge)$
            \State $\text{dfs}(visited,currentSequence,allSequences,G)$  
            \State $\text{visited.remove}(edge.\text{endNode})$
            \State remove the last element from $currentSequence$
        \EndIf
    \EndFor
\EndFunction
\State $allSequences \gets \text{new list}$ 
\ForAll{$startNode$ in $G.\text{nodes}$}
    \State $\text{visited} \gets \text{new set with }startNode$ 
    \State $currentSequence \gets \text{new list}$ 
    \State $\text{dfs}(\text{visited,currentSequence,allSequences,G})$ 
\EndFor
\State \Return $allSequencces$
\end{algorithmic}
\end{algorithm}
This algorithm effectively leverages information from the join graph and employs memoization to swiftly determine a probe order for each data stream that minimizes expected costs. This process takes into account the interdependence among data streams and the costs associated with each probe, ensuring that, despite unforeseen changes in data streams, a more optimal probe order is consistently identified for the multipath flow join process.

\subsection{Statistical Information}
To accurately estimate the cost of data stream probing operations, the operator must first obtain precise historical statistical information regarding the characteristics of the data streams. These statistics primarily include the matching rate between data streams, the average number of records involved in each successful match, and the variation in the size of keys within the data streams. The accurate collection of these statistical indicators is essential for constructing the data stream join cost model.

This paper specifically defines the matching rate $\gamma _{{r_i}}^{{l_i}}$ to quantify the probability of successful matching between the ${l_i}$ data stream and the ${r_i}$ data stream. This probability is calculated as the ratio of the number of successful matches between these two data streams, denoted as $m_{{r_i}}^{{l_i}}$, to the total number of queries initiated, denoted as $q_{{r_i}}^{{l_i}}$, as shown in Eq.~\ref{eq:c9}. Both the number of successful matches and the total number of queries are determined through accumulation. An accurate estimation of the matching rate can assist the operator in understanding the degree of correlation between different data streams, thereby guiding the operator in making more informed decisions regarding probe order during adaptive optimization processes.

\begin{equation}
\label{eq:c9}
\gamma _{{r_i}}^{{l_i}} = \frac{{m_{{r_i}}^{{l_i}}}}{{q_{{r_i}}^{{l_i}}}}
\end{equation}

Furthermore, the operator focuses on the average number of records involved in each successful match $\mu _{{r_i}}^{{l_i}}$. This metric reflects the typical data volume processed during matching. It is calculated by dividing the total number of records accumulated during successful queries, represented as $s_{{r_i}}^{{l_i}}$, by the number of successful matches, indicated as \(m_{{r_i}}^{{l_i}}\). This relationship is illustrated in Eq.~\ref{eq:c10}, where the total number of records accumulated during successful queries is also determined through accumulation.

\begin{equation}
\label{eq:c10}
\mu _{{r_i}}^{{l_i}} = \frac{{s_{{r_i}}^{{l_i}}}}{{m_{{r_i}}^{{l_i}}}}
\end{equation}

Finally, considering that the number of keys in the corresponding state backend of the data stream will change over time, it is particularly important to dynamically update the size of the key count, denoted as ${\kappa _i}$. The operator updates the key count by calculating the impact of the newly added key quantity, $\kappa _i^{{\rm{in}}}$ and the expired key quantity, $\kappa _i^{{\rm{exp}}}$, on the original key size, $\kappa _i^{{\rm{old}}}$, as illustrated in Eq.~\ref{eq:c11}.

\begin{equation}
\label{eq:c11}
{\kappa _i} = \kappa _i^{{\rm{old}}} + \kappa _i^{{\rm{in}}} - \kappa _i^{{\rm{exp}}}
\end{equation}

The operator collects historical statistical information for each data stream during each operating cycle to create a sequence of historical data. To more accurately calculate the cost of each probe order for the upcoming operating cycle, it is essential to predict the statistical information for that cycle. Given that changes in statistical information over the long term are relatively stable and do not exhibit seasonal variations, this solution employs a quadratic exponential smoothing method to forecast the statistical information for the next cycle based on the historical data sequence. Specifically, for the sequences of average record numbers and key quantities, denoted as $\{ {\mu '_1},{\mu '_2},{\mu '_3}, \ldots ,{\mu '_L}\} $ and $\{ {\kappa '_1},{\kappa '_2},{\kappa '_3}, \ldots ,{\kappa '_L}\} $, respectively, Holt's linear trend-based quadratic exponential smoothing is utilized for prediction. Additionally, since the trend of the matching rate evolves over time, damping trend-based quadratic exponential smoothing is applied to predict the sequence of matching rates, denoted as $\{ {\gamma '_1},{\gamma '_2},{\gamma '_3}, \ldots ,{\gamma '_L}\} $.

\section{Experiments}
\subsection{Experimental Environment and Data}
The multi-stream join operator is implemented on the Flink framework, and a Flink Standalone cluster \cite{83} is established under a single-machine configuration, which includes 4 TaskManagers, each configured with 2 slots. All experiments in this chapter are conducted with a parallelism of 8. The hardware and software configuration of the single machine is detailed in Table~\ref{t1}.

\begin{table}[H]
\centering
\caption{Machine Configuration}
\label{t1}
\begin{tabular}{ll}
\hline
Parameter    & Configuration                                      \\ \hline
CPU          & Intel(R) Xeon(R) Platinum 8163 CPU @ 2.50GHz       \\ 
Number of Cores & 96 cores                                          \\ 
Operating System & Alibaba Group Enterprise Linux Server release 7.2 (Paladin) \\ 
Memory       & 503GB                                              \\ 
Hard Disk    & 2.7TB                                              \\ 
Java Version & v1.8.0                                             \\ 
Kafka Version & v2.12-3.0.1                                        \\ 
Zookeeper Version & v3.6.3                                          \\ 
Flink Version & v1.17                                              \\ \hline
\end{tabular}
\end{table}

In the experiments presented in this section, the TPC-DS dataset \cite{84,85,86}, a standard benchmark in the field of big data, is utilized. The version of TPC-DS used is 2.13.0, which includes 17 dimension tables and 7 fact tables, with an average of 18 columns per table. This dataset features both snowflake and star schemas, which are well-suited for real-world query scenarios and optimized for online analytical processing. TPC-DS also provides a data generation tool (TPC-DS tools) that can create datasets of varying sizes based on different scale factors. For example, setting the scale factor to 10 generates a dataset of approximately 10 GB. The following query involves a join of four tables:
\begin{verbatim}
SELECT *
FROM store_returns sr, customer cu, web_returns wr, catalog_returns cr
WHERE cu.c_current_addr_sk = cr.cr_refunded_addr_sk 
AND cu.c_current_addr_sk = sr.sr_addr_sk 
AND cu.c_current_addr_sk = wr.wr_refunded_addr_sk;
\end{verbatim}
In the experiments described in this chapter, a TPC-DS dataset with a scale factor of 10 is utilized. The CSV-formatted data is supplied to the experiment in a random order via Kafka \cite{69} to simulate real-time data streams that change dynamically. The output results are directed to a black hole to prevent I/O operations from influencing the experimental outcomes. The data information for the four tables involved in the aforementioned query from the TPC-DS dataset with a scale factor of 10 is presented in Table~\ref{t2}.

\begin{table}[H]
\centering
\caption{Information of Input Tables}
\label{t2}
\begin{tabular}{llll}
\hline
Table Name       & File Size & Data Size  & Number of Tuples \\ \hline
customer         & 64MB      & 106MB      & 986546           \\ 
catalog\_returns & 212MB     & 331MB      & 2879498          \\ 
store\_returns   & 323MB     & 507MB      & 5750864          \\ 
web\_returns     & 98MB      & 149MB      & 1438434          \\ \hline
\end{tabular}
\end{table}
\subsection{Comparative Experiments}

In this subsection, the runtime of the dynamic adaptive algorithm dpPick is compared to that of various optimization algorithms.

\begin{enumerate}
    \item Default Fixed Initial Order Algorithm (\% fixedOrder): The detection order remains unchanged during runtime.
    \item Selectivity First Algorithm (\% selectivityFirst): Each time, it selects the stream with the minimum selectivity as the next detection target to obtain an optimized detection order. This algorithm is a classical method used for detection order selection in operators like MJoin[35] and GrubJoin[19].
    \item Cost Greedy Algorithm (\% greedy\_MSJ): Each time, it greedily selects the next detection with the minimum cost to obtain an optimized detection order. This is the detection order optimization algorithm used in the MultiStream[42] operator.
\end{enumerate}

\begin{figure}[ht]
\centering
\includegraphics[width=0.6\linewidth]{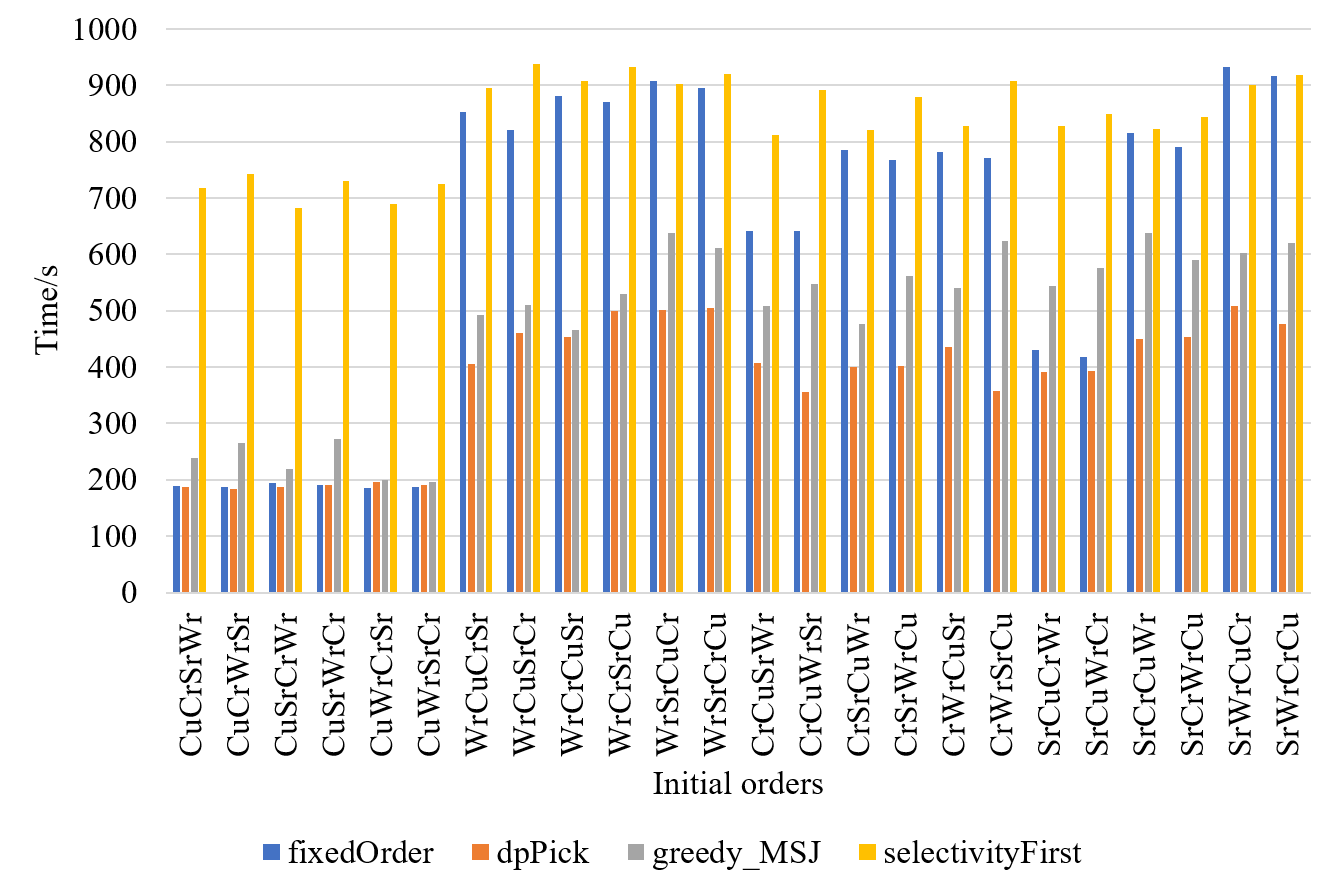}
\caption{Comparative Experiments}
\label{fig:optimization}
\end{figure}

The runtime performance of each algorithm under 24 different initial orders is illustrated in Fig.~\ref{fig:optimization}. In these initial orders, Cu denotes the customer data stream, Cr signifies the catalog\_returns data stream, Wr indicates the web\_return data stream, and Sr represents the store\_returns data stream. For example, when the initial order is CuCrWrSr, the customer data stream is detected first, followed by the catalog\_returns data stream, and then sequentially by the web\_return data streams. Similarly, the detection order of the other data streams adheres to the sequence established by the initial order.

Among the 24 initial orders, the dpPick algorithm achieved the shortest processing time in 22 instances. In the remaining 2 cases, its performance was only slightly inferior to that of the fixedOrder algorithm, with increases of 3.8\% (CuWrCrSr) and 1.0\% (CuWrSrCr), respectively. Compared to the fixedOrder algorithm, the dpPick algorithm reduced runtime by 6.0\% to 53.7\% across the other 22 cases, yielding an average reduction of 31.2\%. In all 24 initial orders, the dpPick algorithm outperformed both the greedy\_MSJ and selectivityFirst algorithms. When compared to the greedy\_MSJ algorithm, the runtime was reduced by 1.5\% to 42.8\%, with an average reduction of 20.4\%. In comparison to the selectivityFirst algorithm, the runtime was decreased by 43.6\% to 75.3\%, with an average reduction of 56.1\%. These results indicate that the dpPick algorithm offers significant advantages in optimizing detection order and highlights its effectiveness in real-time data stream processing. This comparative experiment demonstrates that the dpPick algorithm provides an efficient solution for optimizing runtime in state-based multi-stream join operators during real-time data stream processing, surpassing the default fixed initial order algorithm, the selectivityFirst algorithm, and the cost greedy algorithm.

\subsection{Ablation Experiments}

This section presents comprehensive ablation experiments on the cost model to validate its rationale and assess the effectiveness of the quadratic exponential smoothing prediction method in conjunction with the dpPick algorithm. Three variants of the dpPick algorithm are introduced:
\begin{enumerate}
    \item dpPick\_queryCost algorithm: Considers only the query cost, i.e., sets the matching cost to 0 when calculating the total cost.
    \item dpPick\_matchCost algorithm: Considers only the matching cost, i.e., sets the query cost to 0 when calculating the total cost.
    \item dpPick\_noSmooth algorithm: Does not perform quadratic exponential smoothing on the statistical information and directly uses the statistical information from the previous period for cost calculation.
\end{enumerate}

\begin{figure}[ht]
  \centering
  \includegraphics[width=0.6\linewidth]{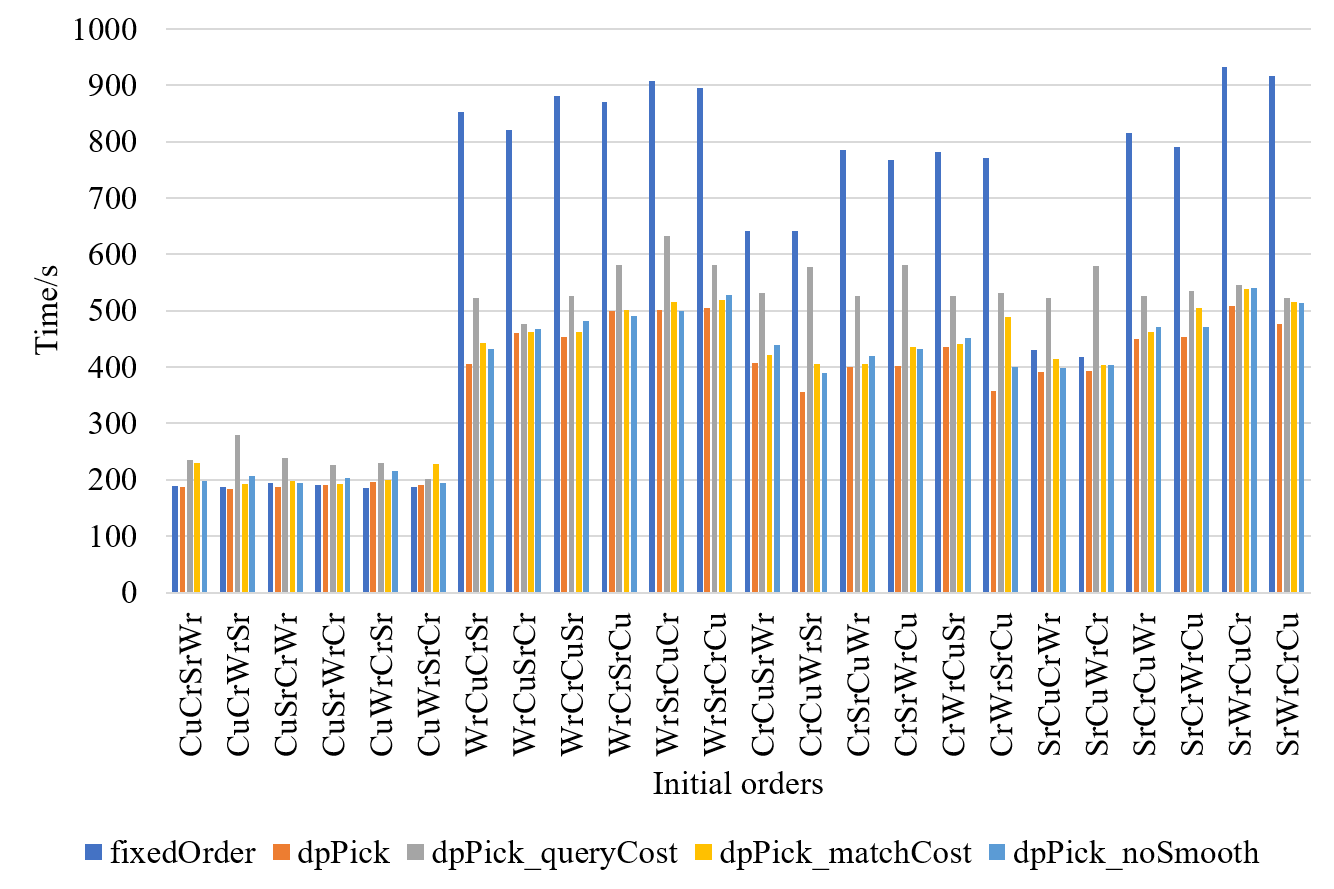}
  \caption{Ablation Experiments}
  \label{fig:melt}
\end{figure}

The results of these three algorithms are compared with those of the dpPick algorithm and the fixedOrder algorithm, as shown in Fig.~\ref{fig:melt}.
When comparing the results of the dpPick\_queryCost and dpPick\_matchCost algorithms to the fixedOrder algorithm, the dpPick\_queryCost algorithm outperforms the fixedOrder algorithm in 16 out of 24 initial orders, while the dpPick\_matchCost algorithm surpasses the fixedOrder algorithm in 18 out of 24 initial orders. This suggests that even when considering only the query cost factor or the matching cost factor individually, both dpPick algorithms can outperform the fixedOrder algorithm.

When comparing the results of the dpPick\_queryCost and dpPick\_matchCost algorithms with the original dpPick algorithm, it is evident that the original dpPick algorithm demonstrates superior performance in all scenarios. The runtime of the dpPick algorithm is reduced by 3.4\% to 32.9\% compared to the dpPick\_queryCost algorithm, with an average reduction of 19.5\%. In comparison to the dpPick\_matchCost algorithm, the runtime of the dpPick algorithm is reduced by 0.4\% to 18.7\%, with an average reduction of 6.3\%. These findings underscore the significance of considering all cost factors and highlight the necessity of the comprehensive cost model proposed in this paper for optimizing the performance of the dpPick algorithm.

Comparing the dpPick\_noSmooth algorithm with the dpPick algorithm, the dpPick algorithm outperforms the dpPick\_noSmooth algorithm in 22 out of 24 initial orders, with performance improvements ranging from 2.5\% to 11.6\%. In the remaining two cases, its performance is lower by 2.0\% (WrCrSrCu) and 0.6\% (WrSrCuCr), respectively. Overall, the dpPick algorithm consistently surpasses the dpPick\_noSmooth algorithm, suggesting that the use of the quadratic exponential smoothing prediction method for statistical forecasting is effective. The instances where the dpPick algorithm performs worse than the dpPick\_noSmooth algorithm are associated with the accuracy limitations of the quadratic exponential smoothing prediction method.

The results of the cost ablation experiments demonstrate that the proposed detection cost model is more adaptable to detection order optimization across various initial detection order scenarios than a single cost factor model, thereby validating the rationale behind the proposed cost model. Additionally, the ablation experiments on prediction methods confirm that employing the quadratic exponential smoothing prediction method for statistical information prediction in the dpPick algorithm is effective and outperforms the approach of using statistical information from the previous period.

\subsection{Influential Factors}

\begin{figure}[htbp]
    \centering
    \begin{minipage}{0.49\linewidth}
        \centering
        \includegraphics[width=0.9\linewidth]{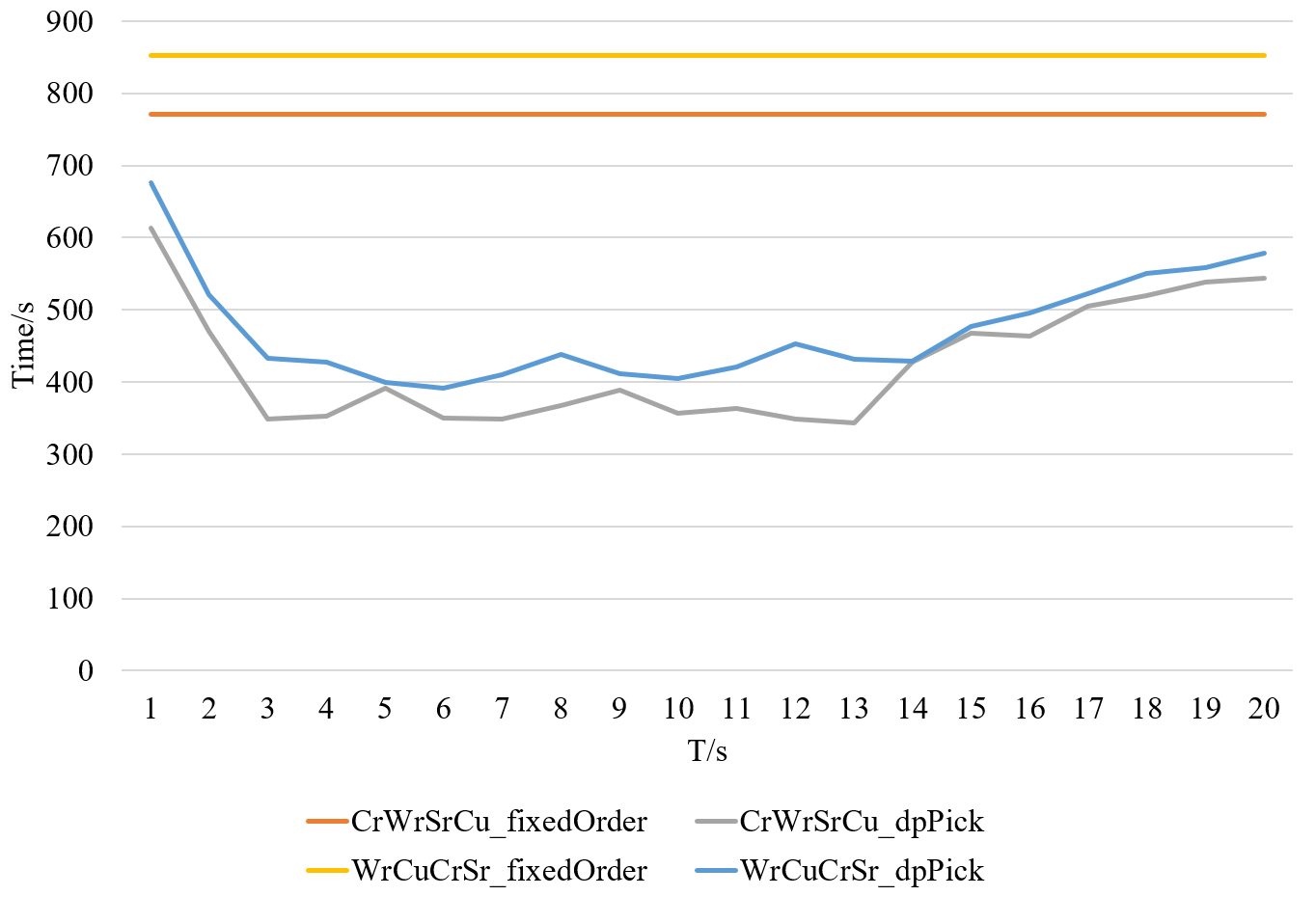}
        \caption{Period of the statistical information sequence}
        \label{period} 
    \end{minipage}
    \begin{minipage}{0.49\linewidth}
        \centering
        \includegraphics[width=0.9\linewidth]{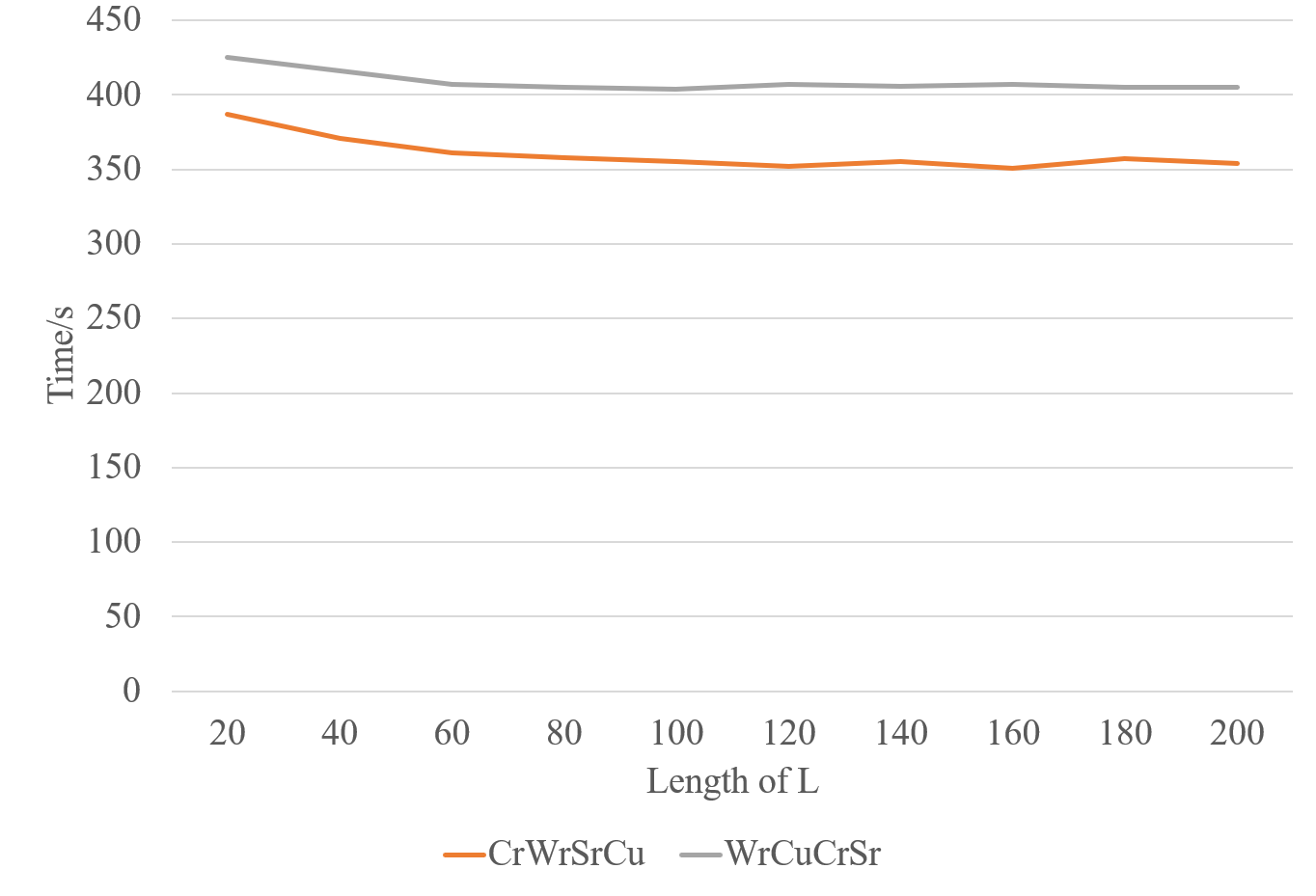}
        \caption{Length of the statistical information sequence}
        \label{length} 
    \end{minipage}
\end{figure}

This section explores and analyzes the impact of the optimization time period \( T \) and the length \( L \) of the statistical information sequence used for prediction on the overall adaptive process. The experiments conducted in this section selected two initial join order scenarios: CrWrSrCu and WrCuCrSr, in which the dpPick algorithm demonstrated the highest efficiency compared to the fixedOrder algorithm. The experimental results are presented in Fig.~\ref{period} and Fig.~\ref{length}.


To investigate the influence of the optimization time period \( T \) on the overall adaptive process, we compare the efficiency performances under various optimization time periods. The experimental results are presented in Fig.~\ref{period}. In this figure, CrWrSrCu\_fixedOrder denotes the performance of the CrWrSrCu initial order using the fixedOrder algorithm as the optimization time period \( T \) varies, while CrWrSrCu\_dpPick indicates the performance of the CrWrSrCu initial order under the dpPick algorithm with changes in the optimization time period \( T \). Similarly, WrCuCrSr\_fixedOrder and WrCuCrSr\_dpPick are defined to represent their respective performances. From the experimental results, it can be observed that the two curves, CrWrSrCu\_dpPick and WrCuCrSr\_dpPick, exhibit similar variations in runtime. When the optimization time period \( T \) is relatively short (1-3 seconds), there is a decreasing trend in runtime. During a moderate optimization period (3-13 seconds), the runtime fluctuates slightly but remains relatively stable. However, when the optimization period is extended (13-20 seconds), there is an overall increasing trend in runtime. In shorter optimization time frames, the limited collection of statistical information may result in suboptimal optimization effects and longer runtimes. As the optimization time period increases, the operator can gather more accurate statistical data, thereby enhancing the precision of cost estimation. With a moderate optimization time frame, the collected statistical information can effectively describe the data for the subsequent period. However, prolonged optimization periods may lead to the collection of outdated information, which can negatively impact the continuity and stability of optimization effects, resulting in increased runtime as the period continues to extend.

To investigate the impact of the length \( L \) of the statistical information sequence used for prediction on the overall adaptive process, we compare the efficiency of performance across different sequence lengths. The experimental results are illustrated in Fig.~\ref{length}. The trends in runtime variations for both cases are consistent. When the sequence length is less than 60, the runtime decreases slightly as the sequence length increases. However, when the sequence length exceeds 60 and goes up to 200, there is no significant change in runtime. A shorter retained statistical information sequence leads to reduced prediction accuracy under the quadratic exponential smoothing prediction method. Conversely, when the statistical information sequence is sufficiently long, prediction accuracy improves. However, an excessively long statistical information sequence has minimal impact on prediction accuracy using the quadratic exponential smoothing method. Given the space and computational overhead associated with retaining and processing statistical information sequences, it is prudent to select a moderate length for the statistical information sequence.

The experimental results indicate that both the optimization time period and the length of the statistical information sequence significantly affect the performance of the adaptive optimization strategy proposed in this study. By establishing suitable optimization time periods and lengths for the statistical information sequence utilized for prediction, the performance of the adaptive optimization strategy presented in this paper can be enhanced, leading to more efficient data processing.

\section{Conclusion}
To address the challenges of adapting multiple stream join operators to unexpected changes in data streams, this paper introduces a runtime-optimized multiple stream join operator. This operator integrates various runtime adaptive optimization strategies into the process of connecting multiple data streams, thereby enhancing the detection sequence. The approach begins by segmenting the operator's operation into distinct time intervals, during which relevant statistical data from the data stream is collected and updated. Historical statistical data is then employed to apply the quadratic exponential smoothing prediction method to forecast the characteristics of the current data stream period. Subsequently, an adaptive optimization algorithm based on a cost model, named dpPick, is developed to refine the detection sequence, improving its adaptability to real-time unknown data streams and increasing operator efficiency. Experimental comparisons using the TPC-DS dataset demonstrate that the proposed method outperforms the fixed initial order method, the selective priority methods of MJoin and GrubJoin operators, and the cost-greedy method of the MultiStream operator in terms of efficiency. Additionally, decomposition experiments on the dpPick algorithm, utilizing cost models and prediction methods, confirmed the foundational principles of the proposed cost model and validated the effectiveness of quadratic exponential smoothing for statistical information. Furthermore, the analysis indicates that establishing appropriate optimization time periods and lengths for statistical information sequences can significantly enhance the overall effectiveness of adaptive optimization strategies. In the future, we plan to utilize more real-world datasets to further validate the effectiveness of this adaptive optimization strategy.

\bibliographystyle{unsrt}
\bibliography{references.bib} 

\begin{thebibliography}{10}

\bibitem{76}
B.~Halstead, Y.~S. Koh, P.~Riddle, et~al.
\newblock Recurring concept memory management in data streams: exploiting data stream concept evolution to improve performance and transparency.
\newblock {\em Data Mining and Knowledge Discovery}, 35(3):796--836, 2021.

\bibitem{77}
M.~Zaki, A.~J. Lee, and P.~K. Chrysanthis.
\newblock Effective access control in shared-operator multi-tenant data stream management systems.
\newblock In {\em Data and Applications Security and Privacy XXXIV}, pages 118--136, Cham, 2020. Springer International Publishing.

\bibitem{78}
S.~Zhang, Y.~Mao, J.~He, et~al.
\newblock Parallelizing intra-window join on multicores: An experimental study.
\newblock In {\em Proceedings of the 2021 International Conference on Management of Data}, pages 2089--2101, Virtual Event China, 2021. Association for Computing Machinery.

\bibitem{79}
H.~Zhang, M.~Qiao, J.~X. Yu, et~al.
\newblock Fast distributed complex join processing.
\newblock In {\em 2021 IEEE 37th International Conference on Data Engineering (ICDE)}, pages 2087--2092, 2021.

\bibitem{80}
Q.~Wang, D.~Zuo, Z.~Zhang, et~al.
\newblock An adaptive non-migrating load-balanced distributed stream window join system.
\newblock {\em The Journal of Supercomputing}, 79(8):8236--8264, 2023.

\bibitem{35}
Viglas~S D, Naughton~J F, and Burger J.
\newblock Maximizing the output rate of multi-way join queries over streaming information sources.
\newblock In {\em Proceedings 2003 VLDB Conference}, pages 285--296, San Francisco, 2003. Morgan Kaufmann.

\bibitem{19}
Gedik B, Wu~K-L, Yu~P S, et~al.
\newblock Grubjoin: An adaptive, multi-way, windowed stream join with time correlation-aware cpu load shedding.
\newblock {\em IEEE Transactions on Knowledge and Data Engineering}, 19(10):1363--1380, 2007.

\bibitem{42}
Manuel D.
\newblock {\em Optimizing Multi-Way Joins for Adaptive, Scale-out Stream Processing}.
\newblock PhD thesis, Rheinland-Pfälzische Technische Universität Kaiserslautern-Landau, Germany, 2023.

\bibitem{17}
Wu~J, Wang Y, Fan X, et~al.
\newblock Toward fast theta-join: A prefiltering and amalgamated partitioning approach.
\newblock {\em Concurrency and Computation: Practice and Experience}, 34(17):e6996, 2022.

\bibitem{18}
Dongdong Zhang, Jianzhong Li, Kimeli K, et~al.
\newblock Slidingwindow based multi-join algorithms over distributed data streams.
\newblock In {\em 22nd International Conference on Data Engineering (ICDE'06)}, pages 139--139, 2006.

\bibitem{20}
Kwon T-H, Lee~K Y, and Kim~M H.
\newblock Load shedding for multi-way stream joins based on arrival order patterns.
\newblock {\em Journal of Intelligent Information Systems}, 37(2):245--265, 2011.

\bibitem{21}
Wang S and Rundensteiner E.
\newblock Scalable stream join processing with expensive predicates: workload distribution and adaptation by time-slicing.
\newblock In {\em Proceedings of the 12th International Conference on Extending Database Technology: Advances in Database Technology}, pages 299--310, 2009.

\bibitem{22}
Fan X, Liu X, Wang Y, et~al.
\newblock Optimizing multi-way theta join for data skew in sub-second stream computing.
\newblock In {\em 2020 IEEE 26th International Conference on Parallel and Distributed Systems (ICPADS)}, pages 476--485, 2020.

\bibitem{23}
Jonathan A, Chandra A, and Weissman J.
\newblock Multi-query optimization in wide-area streaming analytics.
\newblock In {\em Proceedings of the ACM Symposium on Cloud Computing}, pages 412--425, 2018.

\bibitem{24}
Karimov J, Rabl T, and Markl V.
\newblock Astream: Ad-hoc shared stream processing.
\newblock In {\em Proceedings of the 2019 International Conference on Management of Data}, pages 607--622, 2019.

\bibitem{25}
Karimov J, Rabl T, and Markl V.
\newblock Ajoin: ad-hoc stream joins at scale.
\newblock {\em Proceedings of the VLDB Endowment}, 13(4):435--448, 2019.

\bibitem{26}
Le-Phuoc D.
\newblock Adaptive optimisation for continuous multi-way joins over rdf streams.
\newblock In {\em Companion Proceedings of the The Web Conference 2018}, pages 1857--1865, 2018.

\bibitem{27}
Dossinger M and Michel S.
\newblock Optimizing multiple multi-way stream joins.
\newblock In {\em 2021 IEEE 37th International Conference on Data Engineering (ICDE)}, pages 1985--1990, 2021.

\bibitem{28}
Heinrich R, Luthra M, Kornmayer H, et~al.
\newblock Zero-shot cost models for distributed stream processing.
\newblock In {\em Proceedings of the 16th ACM International Conference on Distributed and Event-Based Systems}, pages 85--90, 2022.

\bibitem{29}
Viglas~S D and Naughton~J F.
\newblock Rate-based query optimization for streaming information sources.
\newblock In {\em Proceedings of the 2002 ACM SIGMOD international conference on Management of data}, pages 37--48, 2002.

\bibitem{30}
Vance B and Maier D.
\newblock Rapid bushy join-order optimization with cartesian products.
\newblock {\em ACM SIGMOD Record}, 25(2):35--46, 1996.

\bibitem{31}
Wi~S, Han W-S, Chang C, et~al.
\newblock Towards multi-way join aware optimizer in sap hana.
\newblock {\em Proceedings of the VLDB Endowment}, 13(12):3019--3031, 2020.

\bibitem{32}
Ji~H, Wu~G, Zhao Y, et~al.
\newblock jointree: A novel join-oriented multivariate operator for spatio-temporal data management in flink.
\newblock {\em GeoInformatica}, 27(1):107--132, 2023.

\bibitem{34}
Lai Z, Sun X, Luo Q, et~al.
\newblock Accelerating multi-way joins on the gpu.
\newblock {\em The VLDB Journal}, 31(3):529--553, 2022.

\bibitem{15}
Gomes J and Choi H-A.
\newblock Adaptive optimization of join trees for multi-join queries over sensor streams.
\newblock {\em Information Fusion}, 9(3):412--424, 2008.

\bibitem{16}
Park~H K and Lee~W S.
\newblock Adaptive continuous query reoptimization over data streams.
\newblock {\em IEICE Transactions on Information and Systems}, E92-D(7):1421--1428, 2009.

\bibitem{36}
Zhou Y, Yan Y, Yu~F, et~al.
\newblock Pmjoin: Optimizing distributed multi-way stream joins by stream partitioning.
\newblock In {\em Database Systems for Advanced Applications}, pages 325--341, Berlin, Heidelberg, 2006. Springer.

\bibitem{37}
Babu S, Motwani R, Munagala K, et~al.
\newblock Adaptive ordering of pipelined stream filters.
\newblock In {\em Proceedings of the 2004 ACM SIGMOD International Conference on Management of Data}, pages 407--418, Paris, France, 2004. ACM.

\bibitem{38}
Babu S and Widom J.
\newblock Streamon: an adaptive engine for stream query processing.
\newblock In {\em Proceedings of the 2004 ACM SIGMOD International Conference on Management of Data}, pages 931--932, Paris, France, 2004. ACM.

\bibitem{39}
Golab L and Tamer Özsu M.
\newblock Processing sliding window multi-joins in continuous queries over data streams.
\newblock In {\em Proceedings 2003 VLDB Conference}, pages 500--511, San Francisco, 2003. Morgan Kaufmann.

\bibitem{40}
Lin Q, Ooi~B C, Wang Z, et~al.
\newblock Scalable distributed stream join processing.
\newblock In {\em Proceedings of the 2015 ACM SIGMOD International Conference on Management of Data}, pages 811--825, Melbourne, Victoria, Australia, 2015. ACM.

\bibitem{41}
Dossinger M and Michel S.
\newblock Scaling out multi-way stream joins using optimized, iterative probing.
\newblock In {\em 2019 IEEE International Conference on Big Data (Big Data)}, pages 449--456, Los Angeles, CA, USA, 2019. IEEE.

\bibitem{43}
Yu~S, Zheng Y, Zhang F, et~al.
\newblock Trijoin: A time-efficient and scalable three-way distributed stream join system.
\newblock {\em Journal of Internet Technology}, 24(2):475--485, 2023.

\bibitem{83}
K.~Cai and L.~Ma.
\newblock User behavior data analysis of taobao online based on flink-based k-means algorithm.
\newblock In {\em 2020 International Conference on Applications and Techniques in Cyber Intelligence}, pages 852--859, Cham, 2021. Springer International Publishing.

\bibitem{84}
L.~Liu, H.~Zhang, Y.~Jing, et~al.
\newblock Learned optimizer for online approximate query processing in data exploration.
\newblock {\em IEEE Transactions on Knowledge and Data Engineering}, 2024(1):1--14, 2024.

\bibitem{85}
G.~Chen, T.~Johnson, and M.~Cilimdzic.
\newblock Quantifying cloud data analytic platform scalability with extended tpc-ds benchmark.
\newblock In {\em Performance Evaluation and Benchmarking}, pages 135--150, Cham, 2022. Springer International Publishing.

\bibitem{86}
Y.~Hong, S.~Du, and J.~Leng.
\newblock Evaluating presto and sparksql with tpc-ds.
\newblock In {\em Database Systems for Advanced Applications. DASFAA 2022 International Workshops}, pages 319--329, Cham, 2022. Springer International Publishing.

\bibitem{69}
G.~van Dongen and D.~V.~D. Poel.
\newblock A performance analysis of fault recovery in stream processing frameworks.
\newblock {\em IEEE Access}, 9:93745--93763, 2021.

\end{thebibliography}







\end{document}